\documentclass[preprint,12pt,authoryear]
{elsarticle}

\usepackage{amssymb}
\usepackage{amsmath}

\usepackage{CJKutf8} 

\usepackage{xurl}
\usepackage[unicode]{hyperref}

\makeatletter
\def\ps@pprintTitle{%
   \let\@oddhead\@empty
   \let\@evenhead\@empty
   \let\@oddfoot\@empty
   \let\@evenfoot\@oddfoot}
\makeatother

\begin{document}

\begin{frontmatter}



\title{Advancing credibility and transparency  in brain-to-image reconstruction research: Reanalysis of  Koide-Majima, Nishimoto, and Majima (\textit{Neural Networks}, 2024)
} 


\author[inst1]{Ken Shirakawa}
\author[inst1,inst2]{Yoshihiro Nagano}
\author[inst1,inst2]{Misato Tanaka}
\author[inst1]{Fan L. Cheng}
\author[inst1,inst2,inst3]{Yukiyasu Kamitani}


\affiliation[inst1]{organization={Computational Neuroscience Laboratories, Advanced Telecommunications Research Institute International},
            addressline={Seika-cho}, 
            city={Sorakugun},
            postcode={619-0288}, 
            country={Japan}}

\affiliation[inst2]{organization={Graduate School of Informatics, Kyoto University},
            addressline={Yoshida-honmachi, Sakyo-ku}, 
            city={Kyoto},
            postcode={606-8501}, 
            country={Japan}}

\affiliation[inst3]{organization={Guardian Robot Project, Riken},
            addressline={Seika-cho}, 
            city={Kyoto},
            postcode={619-0288}, 
            country={Japan}}

\begin{abstract}
A recent high-profile study by Koide-Majima et al. (2024) claimed a major advance in reconstructing visual imagery from brain activity using a novel variant of a generative AI-based method. However, our independent reanalysis reveals multiple methodological concerns that raise questions about the validity of their conclusions. Specifically, our evaluation demonstrates that: (1) the reconstruction results are biased by selective reporting of only the best-performing examples at multiple levels; (2) performance is artificially inflated by circular metrics that fail to reflect perceptual accuracy; (3) fair baseline comparisons reveal no discernible advantages of the study's key innovations over existing techniques; (4) the central ``Bayesian'' sampling component is functionally inert, producing outcomes identical to the standard optimization result; and (5) even if the component were successfully implemented, the claims of Bayesian novelty are unsubstantiated, as the proposed method does not leverage the principles of a proper Bayesian framework. These systemic issues necessitate a critical reassessment of the study's contributions. This commentary dissects these deficiencies to underscore the need for greater credibility and transparency in the rapidly advancing field of brain decoding.

\end{abstract}







\end{frontmatter}



\section*{Introduction}
Research at the intersection of neuroscience and artificial intelligence has attracted significant societal and scientific attention, placing a premium on methodological credibility and transparent reporting \citep{macpherson_natural_2021, shirakawa_spurious_2025, kamitani_visual_2025}. A recent high-profile study by \cite{koide-majima_mental_2024} claimed a significant advance in this field, proposing a variant of a generative AI-based method to reconstruct visual imagery from human brain activity using data from \cite{shen_deep_2019}. Their analysis method extended the existing framework \citep{shen_deep_2019} by incorporating Bayesian sampling (stochastic gradient Langevin dynamics, SGLD;\cite{welling_bayesian_2011}) and contrastive language–image pre-training (CLIP)-based semantic features \citep{radford_learning_2021}, reportedly achieving superior performance. The study was widely disseminated through institutional press releases and major media outlets as a breakthrough achievement. More recently, the authors have reframed the study's contribution in subsequent Japanese articles, emphasizing the novelty of the Bayesian method to assert its independence from prior work \citep{JP_Koide_Majima_2024, JP_Majima_2025}. 

In this report, we revisit the findings of \cite{koide-majima_mental_2024} through an independent reanalysis using their publicly released code. Our investigation identifies multiple methodological concerns that raise questions about the study’s conclusions. These issues appear not to be minor mistakes but are indicative of broader systemic problems in the study’s design, analysis, and reporting.

Specifically, our analysis reveals five critical concerns. First, the findings rely on selective reporting at multiple levels, presenting only the most favorable results from a small subset of subjects and samples. Second, the study employs suboptimal quantitative metrics and circular analysis \citep{kriegeskorte_circular_2009}, which do not adequately capture perceptual similarity and risk inflating performance claims. Third, fair baseline comparisons demonstrate that the proposed novel components, Bayesian sampling and semantic features, offer limited improvement in reconstruction quality. Fourth, the core Bayesian sampling procedure is ineffectively implemented, making no clear impact on the final reconstructions. Finally, the conceptual framing of the method as uniquely ``Bayesian'' is a potentially misleading framing, even if the implementation is successful; while prior optimization-based methods like \cite{shen_deep_2019} can also be cast in Bayesian terms, a genuine contribution requires leveraging the framework's strengths, such as uncertainty quantification, which this study fails to do.

In the following section, we substantiate these five concerns with empirical evidence drawn from replication, control analyses, and mathematical investigation. By dissecting this case, our goal is not merely to critique a single study, but to advocate for the establishment of more rigorous and transparent standards for future research in brain decoding.

\section*{Results}

\begin{figure}[!h]
\centering
\includegraphics[width=\linewidth]{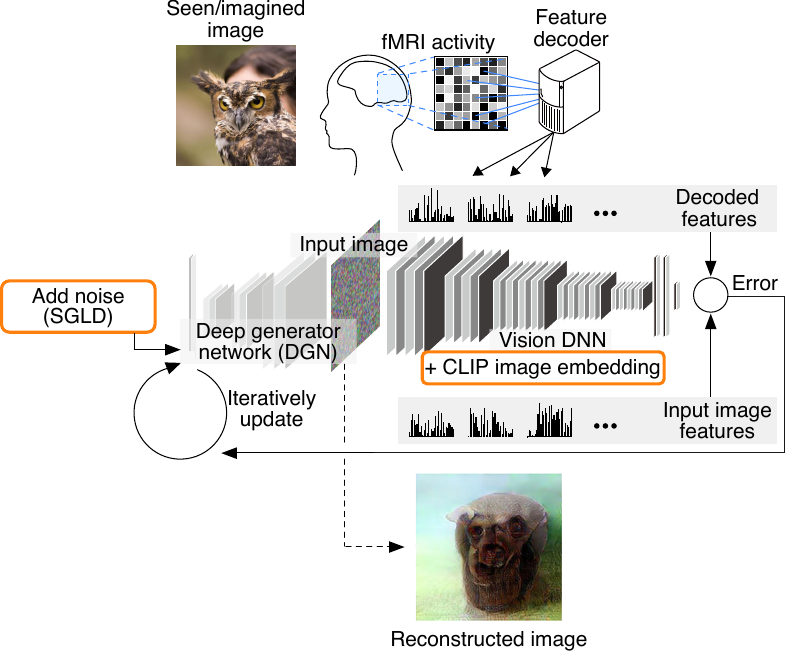}
\caption{\textbf{The iCNN framework in \cite{shen_deep_2019} with the additional components introduced by \cite{koide-majima_mental_2024}.} In the iCNN framework in Shen et al. (2019), an image or the latent vector of a pre-trained deep generator network (DGN) is iteratively updated so that the DNN features of the generated image match those decoded from fMRI activity patterns. The figure shows the additional components proposed by Koide-Majima et al. (2024a), highlighted in orange and superimposed on the original iCNN diagram: (1) the application of stochastic gradient Langevin dynamics (SGLD), which adds Gaussian noise to the latent vector, and (2) the inclusion of CLIP image embeddings as supplementary targets for feature decoding. The overview diagram is adapted from Figure 1 of \cite{shen_deep_2019} under a CC BY 4.0 license.}\label{fig1}
\end{figure}

The method proposed by \cite{koide-majima_mental_2024} is largely based on the iCNN framework introduced by \cite{shen_deep_2019} (Figure~\ref{fig1}). This framework reconstructs images from brain activity in two main steps: feature decoding and image generation. In the feature decoding stage, fMRI responses recorded during visual perception or imagery are translated into deep neural network (DNN) features using linear regression models trained on paired fMRI--feature data. In the image generation stage, an initial latent vector of a pre-trained deep generator network (or a vector of image pixel values) is iteratively updated so that the generated image's DNN features match the decoded ones; the final image is taken as the reconstruction from brain activity. The framework allows multiple configurable options, including the choice of DNN feature layers, the use and type of generator networks, optimization methods, and other processing settings. \cite{shen_deep_2019} presented results using only a few representative combinations of these options.

\cite{koide-majima_mental_2024} highlights two key changes: (1) updating a latent vector via a Bayesian sampling method and (2) introducing the CLIP vision model \citep{radford_learning_2021}. Specifically, they replace the deterministic gradient-based optimization in \cite {shen_deep_2019} with stochastic gradient Langevin dynamics (SGLD) sampling \citep{welling_bayesian_2011}, which is implemented \textit{simply by adding Gaussian noise} to the latent vector at each iteration. In addition, they incorporate CLIP image embedding as an additional target feature alongside conventional vision DNN features (VGG19; \cite{simonyan_very_2015}) to leverage semantic information from the brain. Although not emphasized in their study, their method also differs from Shen’s original method in several technical aspects in the feature decoding procedure, the choice of image generator (DGN), and the formulation of the loss function.

Our analyses are based on the authors’ \ publicly released reconstruction code and decoded DNN features of mental imagery. All analyses were conducted under the ``all'' condition, which refers to the setting where all VGG19 layers are used as decoding targets. This setting was also emphasized in their press release and project page. All reconstructions in this reanalysis were obtained by applying the algorithm \textit{only once} for each sample to avoid cherry-picking from multiple generations, except when the effect of stochastic generation was examined.

\subsection*{1. Multi-hold selective reporting}

A central goal of visual image reconstruction from brain activity is to generate images that faithfully resemble the target perceptual image or mental imagery. Accordingly, the primary evaluation criterion is whether the reconstructed images capture the essential content and overall structure of the intended targets. Qualitative assessment, therefore, plays a crucial initial role in determining reconstruction success, preceding quantitative analyses that typically measure specific aspects of image similarity.

\begin{figure}[!h]
\centering
\includegraphics[width=\linewidth]{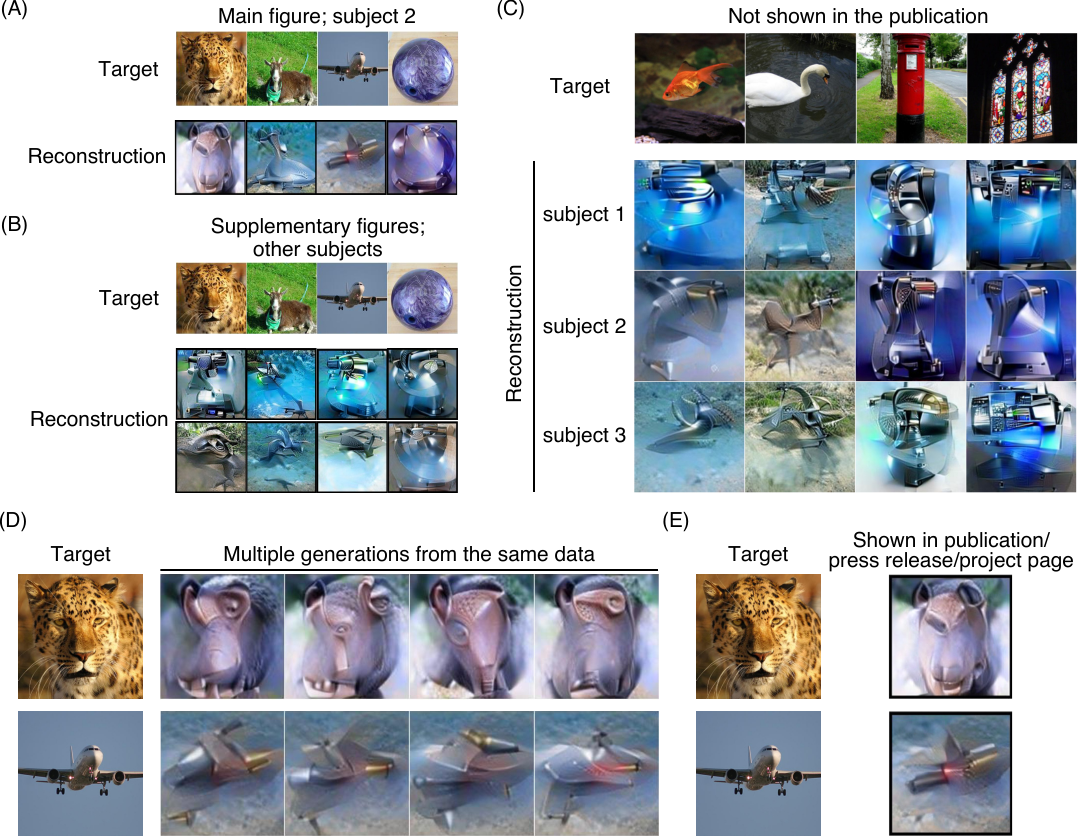}
\caption{\textbf{Reconstruction of mental imagery in \cite{koide-majima_mental_2024}.} (\textbf{A}) Reconstruction results presented in the main figure of  Koide-Majima et al. (2024a). (\textbf{B}) Reconstruction results presented in the supplementary figures of  Koide-Majima et al. (2024a). (\textbf{C}) Reconstruction results for samples not included in their paper. These images are generated using the authors’ publicly released code and data. (\textbf{D}) Reconstruction results obtained by running the reconstruction algorithm multiple times on the same data. (\textbf{E}) Representative reconstruction showcased in the main text and press release of Koide-Majima et al. (2024a). All images (except (\textbf{C}) and (\textbf{D})) are also reproduced from their published figures (CC BY 4.0).
}\label{fig2}
\end{figure}

In \cite{koide-majima_mental_2024}, however, the reconstructions shown in the main figures were limited to a single participant (subject 2; Figure~\ref{fig2}A), while reconstructions from other subjects for the same target images were relegated to the supplementary materials and appeared noticeably lower in quality (Figure~\ref{fig2}B). This selective presentation suggests that results from the best-performing subject were preferentially showcased, without a clear rationale provided for this choice. Clarifying this discrepancy would have been important for transparent evaluation during the peer-review process.

 Selective reporting was also apparent in the choice of target images. Using the publicly available code released by the authors, we generated reconstructions for images that were not shown in the original paper. The resulting outputs, even for subject 2, were qualitatively poorer than those reported (Figure~\ref{fig2}C; see Figure~\ref{figA1} for all results). This discrepancy suggests that the examples presented in the original publication may not reflect typical reconstruction performance.

Further compounding this issue is the fact that their reconstruction algorithm involves multiple sources of randomness. These include not only the sampling process inherent to the Bayesian sampling step, but also random cropping of input images during reconstruction, both of which can introduce variability across trials. In our replication attempts, we observed noticeable differences in the reconstructed images across multiple trials from the same brain data (Figure~\ref{fig2}D). Crucially, despite our considerable effort, we were unable to reproduce results that consistently matched the quality of those showcased in the original paper and press release (Figure~\ref{fig2}E). While some common features were observed, most generated images lacked interpretable facial or object parts characteristic of the showcased examples. This raises the possibility that the showcased reconstructions may have been selectively chosen as the most visually appealing outcomes among a number of attempts. Since the script was released four months after publication, such trial-level selective reporting would have been challenging to detect during the peer review process.

These findings raise concerns that the reconstruction samples presented in the paper may have been selectively chosen, potentially giving a misleading impression that the method reliably captures the key visual features of the target mental images. Without detailed and transparent reporting of the reconstruction sampling procedures and selection criteria, this study leaves open the possibility of selective presentation, raising concerns about scientific transparency.

\subsection*{2. Inappropriate metric and circular evaluation}

\cite{koide-majima_mental_2024} evaluated reconstruction quality using the Inception Score and pairwise image identification accuracy. Although common in related fields, both are insufficient for assessing perceptual similarity between reconstructions and their targets.

The Inception Score measures image naturalness and diversity but ignores correspondence with the true target, allowing high scores for unrelated yet ``realistic'' images. Pairwise identification accuracy, though widely used, can yield inflated scores when category-level distinctions dominate, reflecting coarse rather than fine-grained similarity \citep{shirakawa_spurious_2025}. These metrics thus provide limited evidence of genuine reconstruction fidelity and should only complement qualitative evaluation.

Further compounding the problem is the choice of feature space used to measure similarity in the pairwise identification analysis. The authors evaluated reconstructions using the same feature representation that their algorithm optimized during reconstruction, resulting in a circular analysis \citep{kriegeskorte_circular_2009}. Because the algorithm explicitly drives reconstructions to match those features, testing similarity in the same space merely verifies the algorithm's internal consistency rather than its perceptual validity. Consequently, the reported scores may overstate performance and diverge from actual perceptual similarity.

\begin{figure}[!t]
\centering
\includegraphics[width=\linewidth]{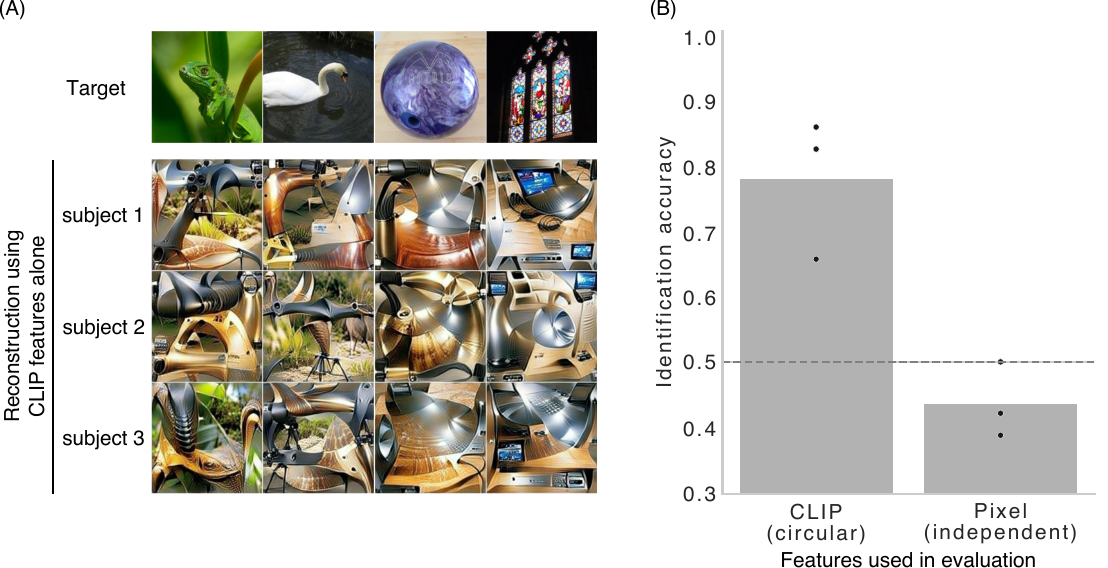}
\caption{\textbf{Effect of circular evaluation.} (\textbf{A}) Reconstructions using only CLIP features decoded from brain activity. Target images were randomly selected from the 10 natural images in the dataset. (\textbf{B}) Pairwise identification accuracy evaluated using different features across 10 natural images. The horizontal axis represents evaluation features; the vertical axis shows identification accuracy. The dashed line indicates the chance level (0.5). Black dots represent individual subjects; gray bars represent the average performance across subjects.
}
\label{fig3}
\end{figure}

To illustrate this circularity, we replicated their analysis using only the decoded CLIP features, omitting the decoded VGG features, which encode essential visual information. Although the resulting reconstructions differed substantially from the targets (Figure~\ref{fig3}), pairwise identification accuracy in the CLIP feature space remained high ($\sim$75\%). In contrast, when evaluated using an independent measure (pixel correlation), accuracy dropped to near chance levels. This demonstrates that when the evaluation metric shares the same feature space as the reconstruction model, identification accuracy no longer reflects the qualitative fidelity of the reconstructions.

Overall, the quantitative evaluations in \cite{koide-majima_mental_2024} do not reliably assess perceptual similarity between reconstructions and their target images. The Inception Score measures image naturalness rather than fidelity, and pairwise identification becomes unreliable when evaluated in a circular feature space. In addition, their comparison with \cite{shen_deep_2019} is ambiguous, as the methodological settings appear to differ from the original Shen et al.'s method reported over 80\% accuracy in human identification tests even for imagery reconstructions, whereas the feature-based identification analysis in  Koide-Majima et al. (2024a) showed substantially lower performance (see \cite{kamitani_visual_2025}, for imagery reconstructions using a recent implementation from our public repository). To clarify these discrepancies, the authors should release all code necessary for full reproducibility. Without human perceptual validation or independent evaluation metrics, their analyses risk overstating reconstruction quality.

\subsection*{3. Insufficiently controlled baselines}

The authors claim that their approach extends the method of \cite{shen_deep_2019}, attributing its effectiveness to two key additions: a Bayesian sampling procedure (SGLD) and the use of semantic features derived from CLIP. However, the effectiveness of these components is not rigorously demonstrated.

First, the supposed benefit of Bayesian sampling is based on an ablation analysis (Fig. 6 in \cite{koide-majima_mental_2024}), which actually tests the impact of the image generator model (prior) rather than the Bayesian sampling procedure itself. As a result, the findings support the utility of incorporating a generator network, an approach already proposed in the original iCNN framework (Fig. 3 in \cite{shen_deep_2019}), but do not provide direct evidence for the added value of Bayesian sampling.

Second, the usefulness of CLIP features is argued based on changes in reconstruction quality when varying their weighting parameter (Fig. 7 in \cite{koide-majima_mental_2024}). The authors claim that removing semantic features lowers identification accuracy while maintaining the Inception Score. However, their results show that identification accuracy remains nearly constant across most parameter settings (except at the default $\lambda_{\textrm{CLIP}}= 0.25$), and that the Inception Score actually improves when CLIP features are excluded ($\lambda_{\textrm{CLIP}}= 0.0$). Given the limitations of both metrics, this discrepancy between the reported data and interpretation provides weak support for the claimed contribution of semantic features to reconstruction quality.

\begin{figure}[!ht]
\centering
\includegraphics[width=0.85\linewidth]{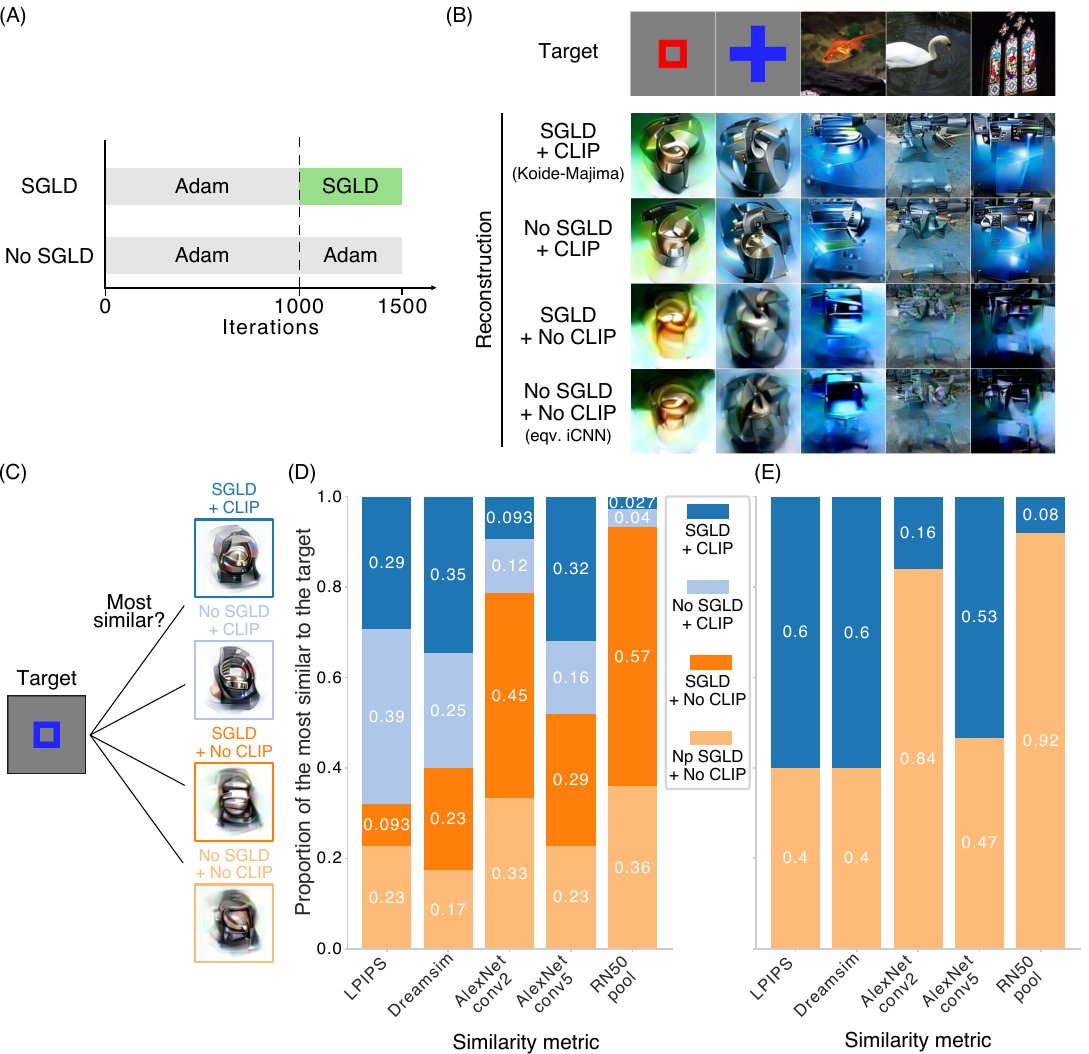}
\caption{\textbf{Controlled ablation analysis of SGLD and CLIP.} (\textbf{A}) Schematic illustration of the updating schedule. Based on the original update schedule in \cite{koide-majima_mental_2024}, the ``SGLD'' condition used 1000 Adam steps followed by 500 SGLD steps, whereas the ``No SGLD'' condition used 1500 Adam steps. (\textbf{B}) Comparison of reconstructed images across different ablation conditions (subject 1). The target images were randomly selected from all 25 target images. The ``SGLD + CLIP'' setting corresponds to  Koide-Majima et al. (2024a), while ``No SGLD + No CLIP'' is essentially equivalent to \cite{shen_deep_2019}. (\textbf{C}) Schematic illustration of the identification analysis. For each target image, reconstructions from each condition are compared to determine which condition yields the result most similar to the target. (\textbf{D}) Identification results across all four conditions. For each similarity metric, the bars show the stacked proportions of samples where each reconstruction condition achieved the highest similarity. AlexNet conv2/5 used correlations at the convolutional layers of AlexNet. RN50 pool used the average of correlations computed across all convolutional layers of OpenAI's ResNet50. (\textbf{E}) Direct comparison between  Koide-Majima et al. (2024a) and the condition equivalent to Shen et al. (2019), formatted as in (\textbf{D}).
}
\label{fig4}
\end{figure}

Third, beyond these two components, the implementation in \cite{koide-majima_mental_2024} differs from the original iCNN framework in several aspects, including the target feature representations, feature decoding procedures, image generator models, and reconstruction loss functions. These modifications, introduced without adequate control analyses, make it difficult to determine whether the reported performance differences can be attributed specifically to SGLD and CLIP. 

To clarify the contributions of the components emphasized by Koide-Majima's study, we conducted an ablation analysis systematically removing the Bayesian sampling procedure (SGLD) and CLIP features, both individually and in combination, yielding four experimental conditions  (Figure~\ref{fig4}A). The reconstruction algorithm in \cite{koide-majima_mental_2024} updates the latent vector using Adam optimization before applying SGLD sampling (Algorithm 5 in \cite{koide-majima_mental_2024}). In the ``No SGLD'' condition, we replaced this step  with Adam-only updates (Figure~\ref{fig4}A). According to their logic, removing both SGLD and CLIP should approximate the original iCNN framework by \cite{shen_deep_2019}, which we therefore used as the baseline for comparison.

We first compared reconstructed image across four conditions (Figure~\ref{fig4}B; see Figures~\ref{figA2}--\ref{figA4} for all subjects). Visual differences among conditions were minor. Removing SGLD had little effect, and adding CLIP features slightly enhanced contrast but did not improve resemblance to the targets.

To quantify reconstruction performance, we conducted an identification analysis comparing reconstructions from each condition with their target images (Figure~\ref{fig4}C). For each target, the reconstruction most similar to the ground truth was identified using metrics independent of the reconstruction algorithm, including perceptual similarity (LPIPS; \cite{zhang2018perceptual}; DreamSim; \cite{fu_dreamsim_2024}) and intermediate feature correlations from DNNs (AlexNet; \cite{krizhevsky_imagenet_2012}; OpenAI’s ResNet \citep{radford_learning_2021}). If SGLD and CLIP truly improved performance, the combined ``SGLD + CLIP'' condition should have consistently outperformed all others. However, this was not observed (Figure~\ref{fig4}D): while slight advantages appeared in some metrics, no condition showed consistent superiority. Even in the direct comparison between the full (``SGLD + CLIP'') and baseline (``No SGLD + No CLIP'') conditions, improvements were minimal (Figure~\ref{fig4}E). These results indicate that the purportedly novel components contribute little to reconstruction quality.

\subsection*{4. Ineffective Bayesian sampling}

\cite{koide-majima_mental_2024} describes their main methodological innovation as reframing image reconstruction as a Bayesian estimation process, using stochastic gradient Langevin dynamics (SGLD) to sample from the posterior distribution over the latent and image spaces. This is implemented by adding a stochastic term to the iterative updates of latent variables in the iCNN algorithm \citep{shen_deep_2019}. Conceptually, this approach could be seen as an extension of the iCNN framework, which itself lends a natural Bayesian interpretation.

However, closer inspection of their SGLD implementation raises concerns about whether true Bayesian sampling meaningfully occurs. As illustrated in Figure~\ref{fig4}A, their algorithm performs 1,000 Adam optimization iterations before only 500 SGLD updates, likely minimizing SGLD's impact. The SGLD step size, defined as $\epsilon_t = \alpha / (\beta + t)^\gamma$ with $\alpha = 0.00015, \beta = 0.15$, and $\gamma = 0.055$, is extremely small (maximum $\approx 0.00016$), several orders of magnitude below typical values used for effective sampling \citep{welling_bayesian_2011}. This greatly limits stochasticity. In addition, the extremely low temperature parameter ($T = 10^{-6}$) further concentrates the likelihood and suppresses exploration, making genuine Bayesian sampling unlikely.

\begin{figure}[!h]
\centering
\includegraphics[width=0.95\linewidth]{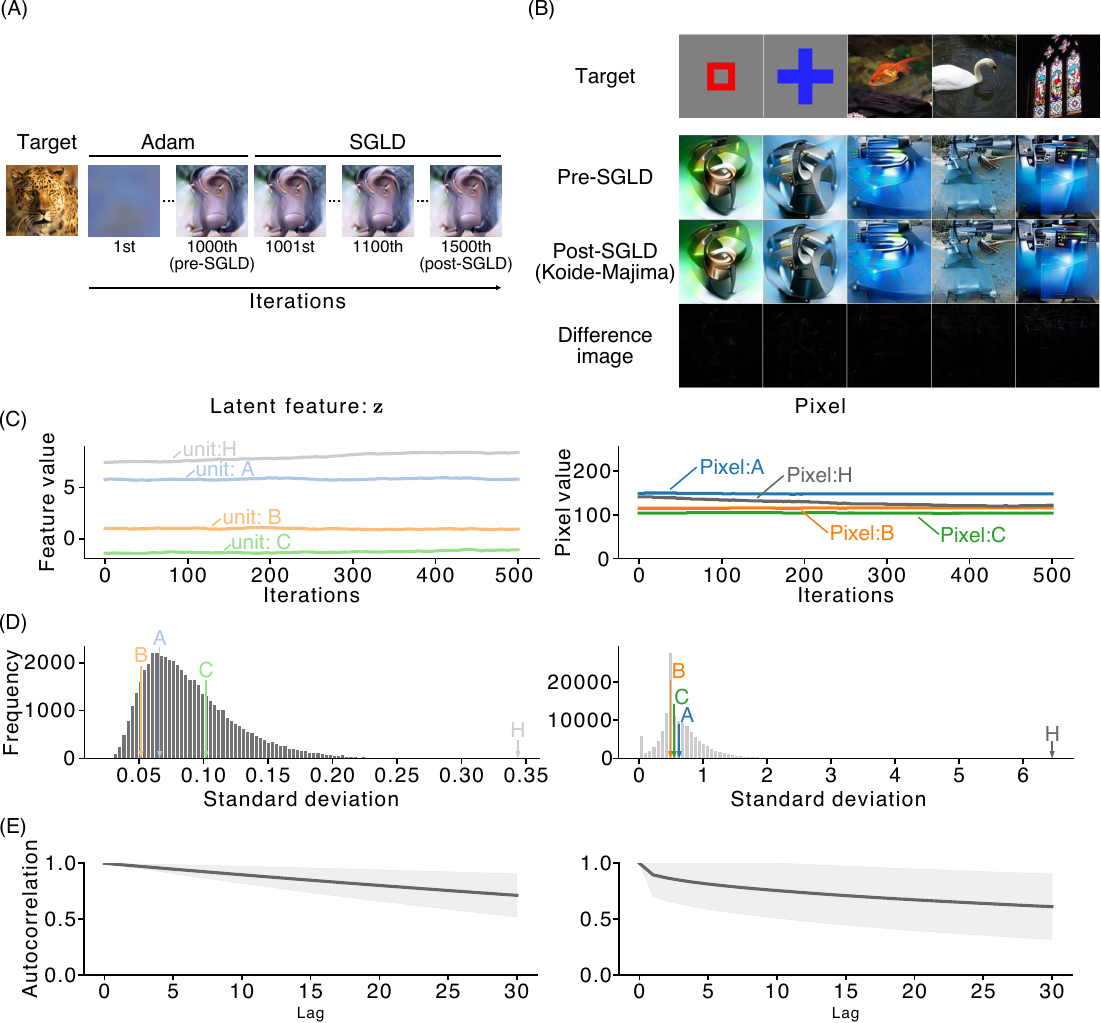}
\caption{\textbf{Lack of substantive effect of SGLD sampling.} (\textbf{A}) Schematic illustration of the updating schedule implemented in \cite{koide-majima_mental_2024}. Their implementation updates the latent vector for 1000 steps using Adam optimization, followed by 500 steps of SGLD sampling.  (\textbf{B}) Comparison of reconstructed images between pre- and post-SGLD. The top row shows the corresponding target images. The middle row shows reconstructions after Adam optimization (pre-SGLD), and the third row shows reconstructions after the SGLD sampling (post-SGLD). The bottom row displays the pixel-wise absolute difference between the pre- and post-SGLD reconstructions, with values ranging from 0 (black) to 255 (white). (\textbf{C}) Value transition of selected latent units (left) or reconstructed image pixels (right) during the SGLD steps shown in (\textbf{A}). Units (or pixels) A-C were randomly selected, while H represents the unit with the largest variation. (\textbf{D}) Histograms of standard deviations of latent units or image pixels during the SGLD period. The values of the units selected in (\textbf{C}) are also shown as a reference. (\textbf{E}) Mean and standard deviation of autocorrelations computed for each latent unit (left) or image pixel (right).
}\label{fig5}
\end{figure}

This two-stage Adam-SGLD process with extremely small hyperparameters suggests that the final reconstructions may be dominated by the preceding Adam optimization. To test this, we compared images obtained immediately after Adam optimization (pre-SGLD) with those produced after the full SGLD process (post-SGLD) (Figure~\ref{fig5}A). The differences were hardly observable (Figure~\ref{fig5}B); the mean pixel-wise mean absolute error (MAE) across three participants and 25 samples was $4.89 \pm 1.00$ (0–255 scale). Increasing the SGLD step size or temperature ($T=1$) did not improve results but rather often degraded structural fidelity (Figure~\ref{figA5}). These findings indicate that SGLD sampling contributes little to reconstruction quality.

We tested whether SGLD effectively samples from the posterior distribution by tracking latent units and image pixels during updates (Figure~\ref{fig5}C). Most units and pixels showed minimal variation, and even the most variable units exhibited only small changes. The distribution of variations confirmed that only 1–2 pixels exhibited notable differences across samples (Figure~\ref{fig5}D). To investigate sampling quality, we computed autocorrelations of both the latent vector and image pixels: effective posterior sampling should produce low autocorrelation across steps \citep{roy_convergence_2020}. However, autocorrelations remained consistently high (Figure~\ref{fig5}E), indicating strong temporal dependency and insufficient exploration of the posterior. Here, we treated the initial 1000 Adam updates as the burn-in phase and analyzed the subsequent 500 SGLD steps. For completeness, we discarded the initial 250 steps of the SGLD sampling as an additional burn-in, but similar results were obtained.

Together, these findings indicate that the SGLD procedure in Koide-Majima et al.'s framework contributes little to both reconstruction quality and posterior exploration. In its current form, the SGLD component appears to have a negligible effect, calling into question the central claim that Bayesian sampling is critical to their results. 

\subsection*{5. Misleading Bayesian framing}

In their subsequent Japanese review articles and institutional press releases \citep{JP_Majima_2025, qst_nict_osaka_jst_press_2023}, Koide-Majima and colleagues have presented their approach as a distinct ``Bayesian'' framework, emphasizing its independence from the earlier iCNN method. However, as shown in the previous section, their implementation of SGLD fails to operate effectively, yielding results equivalent to iCNN. Even if it were properly implemented, a deeper question remains: does their framework offer any meaningful methodological innovation as a Bayesian approach beyond nominal framing? In this section, we first review the methodological structure of their framework and then critically examine the validity of their claim.

In their formulation, \cite{koide-majima_mental_2024} aimed to estimate the posterior distribution of an image $\mathbf{s}$ given the DNN features decoded from brain activity $\hat{\boldsymbol{\phi}}$.  Using the input variable $\mathbf{z}$ to the deep generator network as a latent variable, the posterior distribution of images is decomposed as:
$$
p(\mathbf{s} \mid \hat{\boldsymbol{\phi}}) 
= \int p(\mathbf{s} \mid \mathbf{z}) \, p(\mathbf{z} \mid \hat{\boldsymbol{\phi}}) \, d\mathbf{z}.
$$

Here,  $p(\mathbf{s} \mid \mathbf{z})$ is defined by the deterministic mapping of the deep generator network  $\mathbf{s} = g(\mathbf{z})$. SGLD is then used to sample from the latent posterior $p(\mathbf{\mathbf{z}} \mid \hat{\boldsymbol{\phi}})$ by iteratively updating as

$$
\mathbf{z}_{t+1} = \mathbf{z}_t + \frac{\epsilon_t}{2} \nabla_{\mathbf{z}} \ln p(\mathbf{z}_t \mid \hat{\boldsymbol{\phi}}) + \boldsymbol{\eta}_t,
$$

where $\boldsymbol{\eta}_t$ represents Gaussian noise sampled from $\mathcal{N}(0, \epsilon_t I)$. The latent posterior $\ln p(\mathbf{z}_t \mid \hat{\boldsymbol{\phi}})$ is modeled through a log-likelihood function
$$
\ln p(\hat{\boldsymbol{\phi}} \mid \mathbf{z}) \propto -L(f(g(\mathbf{z})), \hat{\boldsymbol{\phi}}),
$$
assuming a non-informative prior $p(\mathbf{z})$. $L$ denotes a feature-matching loss between the decoded DNN features and those extracted from the generated image, where $f$ represents a feature extraction network such as VGG19 or CLIP. This process drives the generated image features to align with the decoded ones. Notably, if the Gaussian noise term ($\boldsymbol{\eta}_t$) is removed, the updating algorithm reduces to the iCNN optimization procedure. As shown in the previous section, this noise term appears to be ineffective in their implementation.

If properly implemented, the iterative process of SGLD would perform sampling from the posterior distribution without requiring explicit characterization of the posterior distribution itself, such as finding the most probable point, or the maximum a posteriori (MAP) estimate. In particular, obtaining the MAP estimate over the image space $\mathbf{s}_\text{MAP}$  is nontrivial, even if a MAP estimate in the latent space $\mathbf{z}_\text{MAP}$ is calculated, for instance, through deterministic optimization as in iCNN. This difficulty arises because the nonlinearities and potentially complex Jacobian structures of the deep generator network could cause misalignment between the modes of the latent and image spaces, meaning that  $\mathbf{s}_\text{MAP} = g(\mathbf{z}_\text{MAP})$ does not generally hold. While Koide-Majima et al.’s approach does not directly resolve this issue, it can still, in principle, generate samples from the posterior distribution in the image space via the samples in the latent space: $\mathbf{s}_t = g(\mathbf{z}_t)$.

Although such samples could enable richer uncertainty quantification and more robust posterior estimation, such potential is not demonstrated in their study. The implementation uses only the final sample $\mathbf{s}_T = g(\mathbf{z}_T)$ for reconstruction, effectively collapsing the Bayesian sampling into a single-point estimate without leveraging uncertainty information. If the goal is to obtain a point estimate, more principled Bayesian approaches exist. For example, the posterior mean can be approximated by averaging multiple samples
$$  
\mathbf{s}_{\text{PM}} = \mathbb{E}[g(\mathbf{z})] \approx \frac{1}{T} \sum_{i=1}^{T} g(\mathbf{z}_i),  
$$
yielding a more stable estimate than a single sample, though requiring greater computational cost.

It should also be noted that many optimization-based methods can also be interpreted as Bayesian point estimation under certain assumptions. For instance, ridge regression corresponds to maximum a posteriori (MAP) estimation under Gaussian priors \citep{bishop_pattern_2006}. Similarly, the iCNN method can be viewed as a MAP estimation defined on the generator manifold (see \ref{app2}). Thus, the distinction between “Bayesian” and “non-Bayesian” is largely interpretive. The interpretability as Bayesian does not, by itself, convey any practical advantage.

Together, although \cite{koide-majima_mental_2024}’s method adopts Bayesian terminology through SGLD sampling, it produces only a single reconstruction that closely matches results from optimization-based methods. In practice, it behaves as stochastic optimization rather than true Bayesian inference, even if properly implemented. A genuine Bayesian approach would exploit the full posterior distribution to quantify uncertainty and capture the inherent ambiguity of decoding results.

\section*{Discussion}

Our independent reanalysis of \cite{koide-majima_mental_2024} reveals multiple interconnected methodological concerns that critically undermine the validity of their central claims regarding successful reconstruction of mental imagery from brain activity. While the study has been widely disseminated through institutional press releases and media coverage, the empirical evidence obtained from systematic replication and controlled analyses diverges substantially from the claims presented in the paper and associated communications.

Our analysis reveals five key issues: selective reporting of only the most favorable results gives a misleading view of performance; quantitative evaluations are compromised by circularity, as the same features are used for both optimization and evaluation; claimed methodological innovations (SGLD-based Bayesian sampling and CLIP features) provide no consistent improvement over standard baselines when fairly compared; SGLD sampling is ineffectively implemented, producing reconstructions nearly identical to deterministic optimization; and finally, the ``Bayesian'' framing is largely nominal and of limited practical value. The approach does not realize core Bayesian advantages such as uncertainty quantification, and in fact, the existing iCNN framework can also be interpreted in Bayesian terms.

Beyond these empirical and theoretical limitations, concerns regarding transparency and reproducibility warrant particular attention. Crucially, at the time of peer review, the exact procedure for evaluation was not described in the paper and neither the code nor data were made available, making it impossible for reviewers to assess essential aspects of the methodology. As a result, reviewers could not identify certain issues, such as the circularity introduced by using the features both for optimization and for evaluation, an aspect only revealed through later inspection of the released code. Similarly, the ineffective implementation of SGLD-based sampling, which results in nearly deterministic outputs rather than genuine sampling from the posterior, was not apparent from the manuscript alone and became clear only with access to implementation details. Thus, the study may have been published without the level of scrutiny and methodological assessment that would have been possible if code and full methodological details had been transparently provided. Although partial code was released four months after publication, essential components including scripts for reproducing the results reported in the figures, are still unavailable. This incomplete disclosure prevents full reproducibility and, combined with the substantial discrepancies between the selectively presented reconstructions in the paper and those obtainable from the public code, raises questions about the representativeness of the reported results.

The broader context of the study's promotion merits attention. Despite the methodological concerns identified here, the study was extensively disseminated through joint press releases issued by the National Institutes for Quantum Science and Technology (QST), the National Institute of Information and Communications Technology (NICT), Osaka University, and the Japan Science and Technology Agency (JST) \citep{qst_nict_osaka_jst_press_2023}, as well as through major media outlets (e.g., \cite{nikkei_inc_nictjstai_2023, mainichi_2023, kyodo_2023, jiji_2023}. In these communications, selectively chosen reconstructions were presented as representative achievements in mental image reconstruction. Furthermore, while the ``Bayesian'' framing received limited emphasis in the original paper, it was amplified in subsequent press releases and Japanese review and tutorial articles \citep{JP_Koide_Majima_2024, JP_Majima_2025} in ways that may obscure the continuity with prior work \citep{shen_deep_2019} and overstate the methodological novelty of the contribution.

The amplification of claims that lack sufficient validation has far-reaching consequences well beyond the scope of a single study. For the general public, it can fuel unrealistic expectations about the present state of brain decoding technology, only to erode trust in science when those expectations go unmet. Within the scientific community, such publicity risks skewing resource distribution and recognition, as attention is diverted to headline-grabbing yet insufficiently validated results at the expense of rigorous, reproducible work that may not receive equal visibility. The effects on early-career researchers are especially worrying: facing hyped claims and seemingly unattainable standards, some may become disillusioned or discouraged from entering or continuing in the field, while others may wrongly blame themselves for being unable to reproduce overstated findings--misattributing systemic flaws to personal inadequacy. Altogether, this environment can foster misconceptions about what genuine progress in brain decoding research looks like and hinder healthy, sustainable advancement in the field.

The issues identified in this study exemplify challenges that extend beyond a single publication and reflect broader systemic issues in how complex computational neuroscience research is conducted, evaluated, and communicated. Our reanalysis underscores the critical importance of several fundamental research practices that are essential for maintaining scientific credibility: (1) presenting results in a balanced and representative manner that accurately reflects typical performance rather than selectively emphasizing the most favorable outcomes, (2) validating methods using evaluation metrics that are independent of the optimization objective and employing sufficiently controlled baselines that isolate the contribution of specific methodological components, (3) releasing complete and well-documented code that enables full reproduction of all results reported in the paper, including both visualization and quantitative evaluation, (4) ensuring that research claims in both papers and press releases accurately reflect the strength of the evidence and acknowledge limitations transparently, and (5) facilitating timely independent verification by the broader research community through prompt release of code and data. Reaffirming these practices is particularly important as research at the intersection of neuroscience and artificial intelligence grows increasingly complex, computationally intensive, and attracts substantial public attention and media coverage.

At the same time, it is important to recognize that research in this field, like all scientific inquiry, is inherently exploratory and subject to unexpected challenges. Methodological limitations and errors are often an unavoidable part of the process. What distinguishes credible science is not the absence of such issues, but their open acknowledgment and a genuine willingness to address them constructively. When problems or weaknesses are identified, the responsible course is to evaluate the evidence carefully, discuss limitations transparently, and correct errors as needed. These responses reinforce, rather than diminish, scientific credibility. By cultivating a research culture that prioritizes rigor, transparency, and reproducibility alongside innovation and impact, we can strengthen the foundations of basic neuroscience, deepening our understanding of how mental experiences are represented in the brain, and support the responsible development of practical neurotechnologies. This approach not only maintains public trust but also accelerates real progress in brain decoding research, both at the fundamental and applied levels.

\section*{Acknowledgments}
We thank our laboratory team for their invaluable feedback and insightful suggestions on the manuscript.
This work was supported by the Japan Society for the Promotion of Science (JSPS: KAKENHI grant, 20H05954, JP25H00450).

\section*{Code availability}
The code used for this analysis is available here (\url{https://github.com/KamitaniLab/repro_mental_image_recon}). To minimize unintended discrepancies across different analysts, most of the code directly reuses functions from Koide-Majima's Github repository (\url{https://github.com/nkmjm/mental_img_recon}) and procedures implemented in their demonstration code (\url{https://colab.research.google.com/drive/1gaMoae0ntiT94-rQUMymkZboNc-imTzl?usp=drive_link}). Note that this original repository contains multiple sources of randomness without proper control of random seeds. Despite our extensive efforts, we were also unable to fully control this randomness in our reproduction. Thus, please understand that running these codes will not reproduce exactly the same results as those reported in this analysis.





\clearpage
\appendix
\renewcommand{\thefigure}{A\arabic{figure}}
\setcounter{figure}{0}
\section{Supplementary figures}
\label{app1}

\begin{figure}[!h]
\centering
\includegraphics[width=1.0\linewidth]{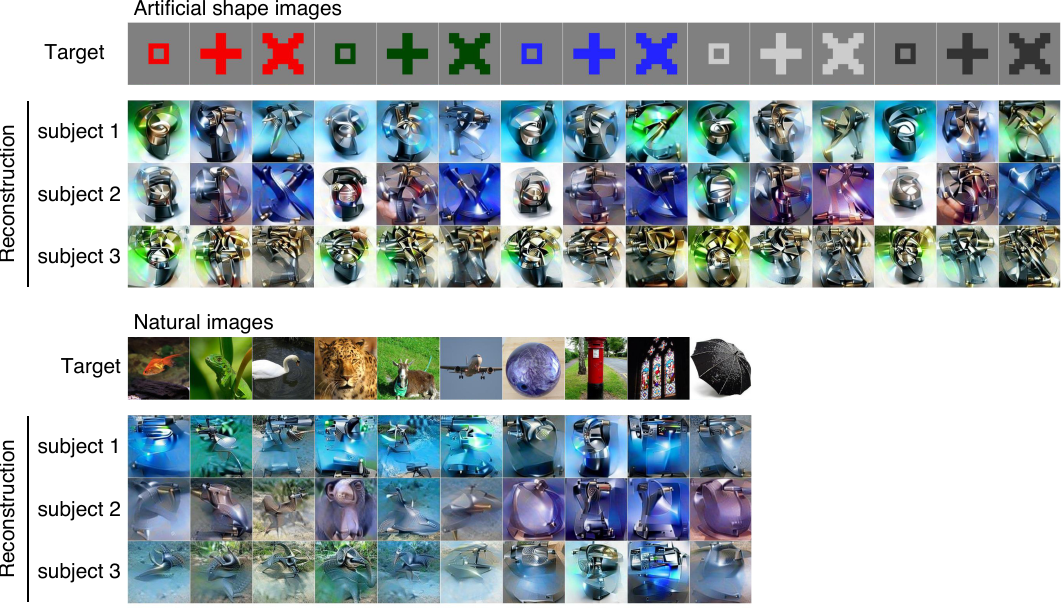}
\caption{\textbf{All results of the reconstruction of mental imagery from brain activity.} Reconstruction results for all target images used in our replication analysis, including artificial and natural images. Each row corresponds to one subject (subjects 1–3), and each column to one target image.}
\label{figA1}
\end{figure}

\begin{figure}[!t]
\centering
\includegraphics[width=1.0\linewidth]{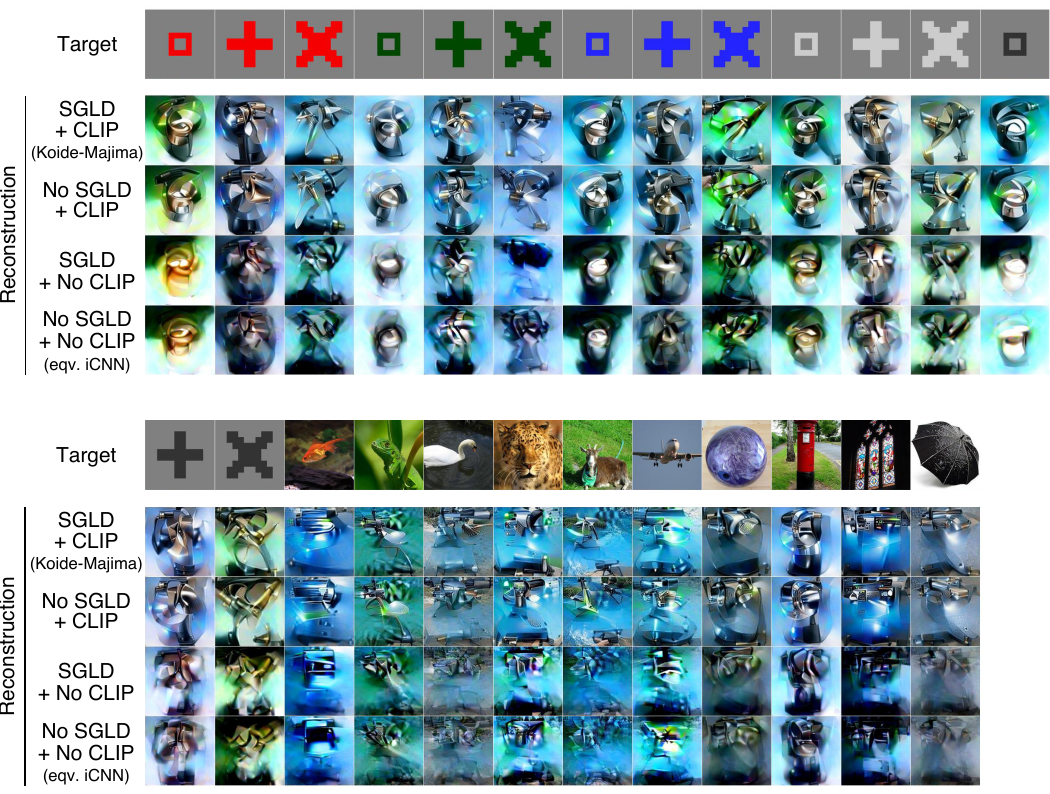}
\caption{\textbf{All reconstruction results of ablation analysis (subject 1).} Reconstructions from four conditions are shown: the full method proposed by Koide-Majima et al. (``SGLD + CLIP''), and three ablated variants in which the Bayesian sampling component (``No SGLD + CLIP''), the semantic component (``SGLD + No CLIP''), or both were removed (``No SGLD + No CLIP'').
}
\label{figA2}
\end{figure}

\begin{figure}[!t]
\centering
\includegraphics[width=1.0\linewidth]{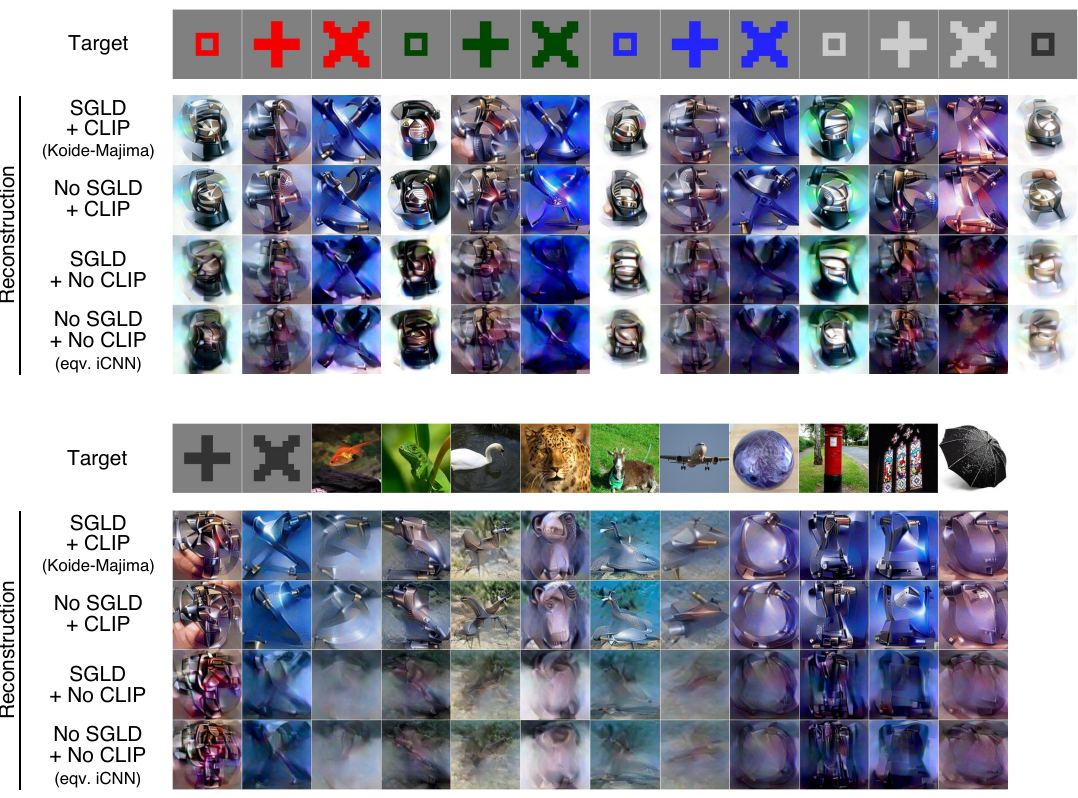}
\caption{\textbf{All reconstruction results of ablation analysis (subject 2).} Reconstructions from four conditions are shown: the full method proposed by Koide-Majima et al. (``SGLD + CLIP''), and three ablated variants in which the Bayesian sampling component (``No SGLD + CLIP''), the semantic component (``SGLD + No CLIP''), or both were removed (``No SGLD + No CLIP'').
}\label{figA3}
\end{figure}

\begin{figure}[t]
\centering
\includegraphics[width=1.0\linewidth]{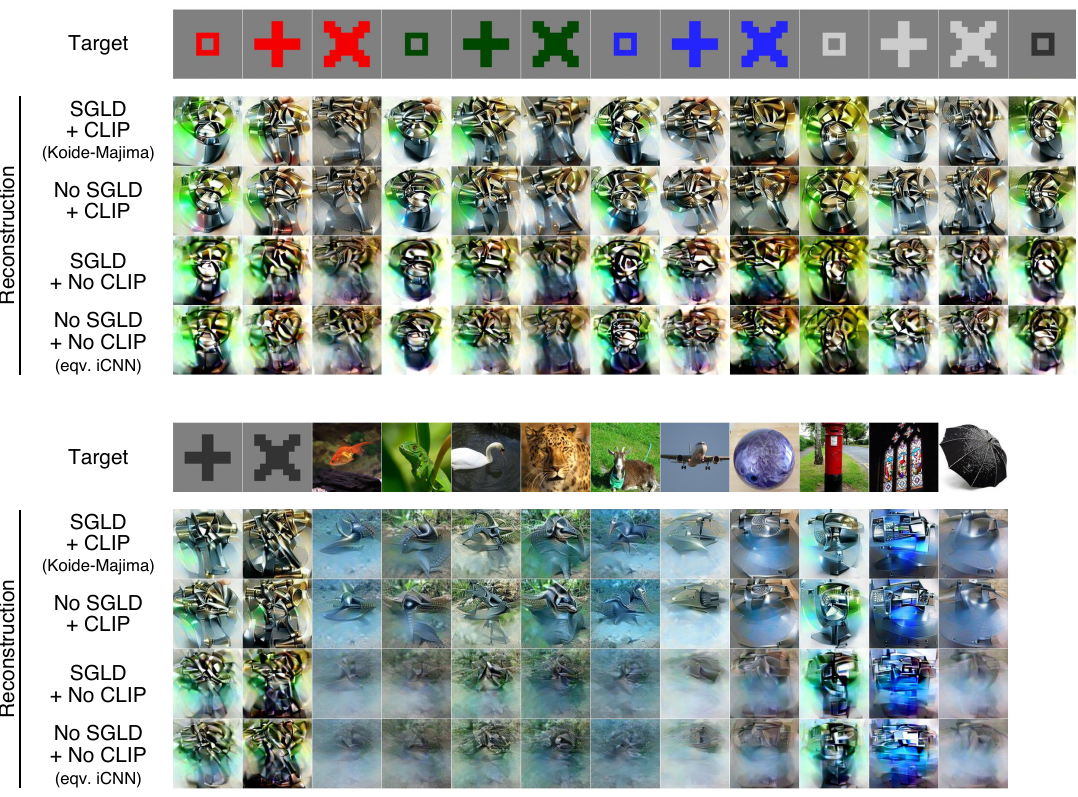}
\caption{\textbf{All reconstruction results of ablation analysis (subject 3).} Reconstructions from four conditions are shown: the full method proposed by Koide-Majima et al. (``SGLD + CLIP''), and three ablated variants in which the Bayesian sampling component (``No SGLD + CLIP''), the semantic component (``SGLD + No CLIP''), or both were removed (``No SGLD + No CLIP'').
}
\label{figA4}
\end{figure}

\begin{figure}[t]
\centering
\includegraphics[width=1.0\linewidth]{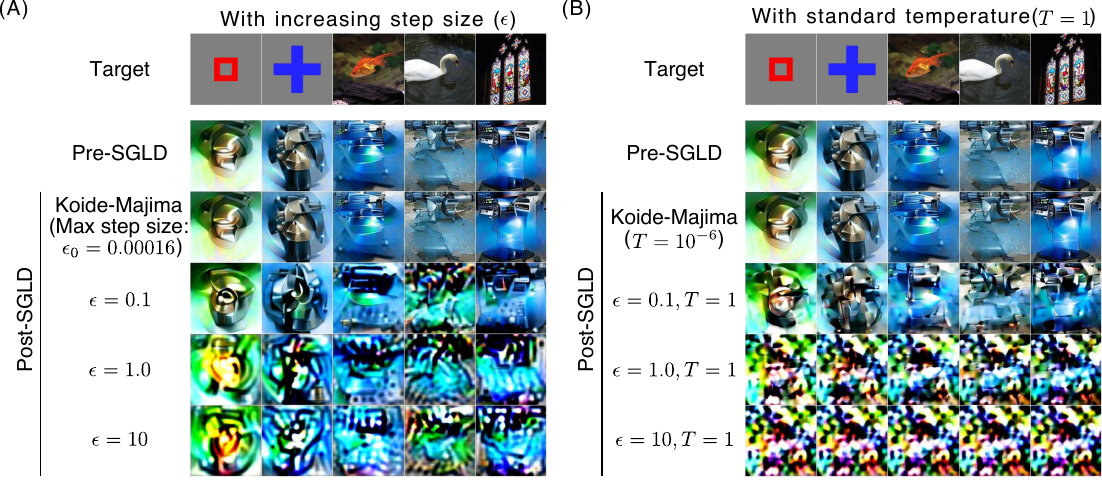}
\caption{\textbf{Comparison of reconstructed images between pre- and post-SGLD under different hyperparameter settings.} (\textbf{A}) Reconstructions with increasing step size. In the original implementation by Koide-Majima, the maximum step size was set to $\epsilon_0 = 0.00016$, and decreased over the SGLD sampling schedule. Here, we instead fixed the step size to higher values ($\epsilon =0.1, 1.0, 10$) across all sampling steps. These post-SGLD conditions were initialized from the same latent vector of the pre-SGLD condition. (\textbf{B}) Reconstructions with standard temperature ($T=1$). In the Koide-Majima implementation, an unusually low temperature parameter ($T=10^{-6}$) was used in the likelihood modeling. This temperature parameter controls the stochasticity of sampling, with smaller values leading to more deterministic outputs. In general, $T=1$ is considered a standard setting. Here we set $T=1$ and varied the step size ($\epsilon=0.1, 1.0, 10$) during the SGLD phase. These post-SGLD conditions were initialized from the same latent vector of the pre-SGLD condition.}
\label{figA5}
\end{figure}




\clearpage
\section{Bayesian interpretation of iCNN}
\label{app2}
This section demonstrates how optimization-based machine learning methods can be reinterpreted as maximum a posteriori (MAP) estimation under appropriately chosen probabilistic models.
Our goal is to show that iCNN, which appears to be a purely optimization-based method, can be understood as performing MAP estimation on the manifold spanned by a generative model.
To build this understanding systematically, we begin by introducing foundational concepts: change of variables for random variables and the role of Jacobian determinants in probability transformations.
We then illustrate the MAP interpretation framework through ridge linear regression as a pedagogical warm-up example, before applying the same interpretive lens to iCNN.
The key insight we develop is that a carefully chosen Jacobian-based prior distribution enables this reinterpretation, revealing a connection between optimization and probabilistic inference.

\subsection*{Preliminary: Change of variables and Jacobian}
In general, when a continuous random variable $\mathbf{s}$ is obtained from a different random variable $\bf{z}$ through a deterministic transformation $\mathbf{s} = g({\mathbf{z}})$, the probability density $p_S({\mathbf{s}})$ of the random variable $\mathbf{s}$ can be expressed using the probability density $p_Z(\mathbf{z})$ of the pre-transformation random variable $\mathbf{z}$ as follows:
\begin{equation}
    p_S({\mathbf{s}}) = p_Z(g^{-1}({\mathbf{s}})) \left| \det J_{g^{-1}}({\mathbf{s}}) \right|. \label{eq:change-of-variables}
\end{equation}
Note that this transformation holds when the $g$ is invertible and smooth, where $J_{g^{-1}}({\mathbf s}) = \partial g^{-1} / \partial {\mathbf{s}}$ is the Jacobian of $g^{-1}$, and $| \det J_{g^{-1}}({\mathbf{s}}) |$ represents its determinant. Modeling intractable probability distributions using this change of variables has attracted considerable attention in recent years and is studied as normalizing flows \citep{tabak_density_2010, tabak_family_2013, rezende_variational_2015}. In normalizing flows, intractable probability distributions of observations $p_S({\mathbf s})$ such as images are modeled via invertible neural networks $g$ with explicit latent probability distributions such as Gaussian distributions.

In \autoref{eq:change-of-variables}, the matrix determinant $\left| \det J_{g^{-1}}({\bf s}) \right|$ is generally interpreted as the infinitesimal volume change induced by applying the nonlinear transformation $g$ at the point of interest $\mathbf{s}$. Due to the existence of such infinitesimal volume changes, MAP estimates computed in the pre- and post-transformation spaces do not necessarily coincide:
\begin{equation}
    \mathbf{s}_{\mathrm{MAP}} = \arg\max_{\mathbf{s}} p_S(\mathbf{s}) \neq g\left( \arg\max_{\mathbf{z}} p_Z(\mathbf{z}) \right) = g(\mathbf{z}_{\mathrm{MAP}}). \label{eq:map-discrepancy}
\end{equation}
Such discrepancy does not occur when the transformation $g$ is linear or when the Jacobian determinant is constant regardless of location, but becomes pronounced for highly nonlinear transformations such as deep neural networks.
As a concrete example, considering a random scalar variable $s$ following a one-dimensional log-normal distribution allows an intuitive understanding of the MAP estimate discrepancy. Specifically, let $z \sim \mathcal{N}(\mu, \sigma^2)$ and define $s = g(z) = e^z$. Then, by the change of variables formula, $p_S(s) = p_Z(\ln s) \cdot |J_{g^{-1}}(s)|$ where $J_{g^{-1}}(s) = 1/s$. Therefore, $\ln p_S(s) = \ln p_Z(\ln s) - \ln s$, and maximizing this yields $s_{\mathrm{MAP}} = e^{\mu - \sigma^2} \neq e^{\mu} = g(z_{\mathrm{MAP}})$.
Since the Gaussian distribution has a bell-shaped form centered at the expectation $\mu$, the mode after exponential transformation would be at $e^{\mu}$, yet due to the Jacobian effect stretching the probability density, the peak of the log-normal distribution shifts to $e^{\mu - \sigma^2}$.

\subsection*{Interpreting iCNN as MAP estimation on generator manifold}
To illustrate how optimization problems can be reinterpreted as MAP estimation, we first consider ridge linear regression as a well-understood example before tackling the more complex case of iCNN.
Ridge linear regression estimates parameters $\mathbf{w} \in \mathbb{R}^d$ for input data $X \in \mathbb{R}^{n \times d}$ and output data $\mathbf{y} \in \mathbb{R}^n$ by solving the following optimization problem:
\begin{equation}
    \mathbf{w}^* = \arg\min_{\mathbf{w}} \| X \mathbf{w} - \mathbf{y} \|_2^2 + \lambda \| \mathbf{w} \|_2^2. \label{eq:ridge-regression}
\end{equation}
Here, $\lambda > 0$ is a regularization parameter.
The standard interpretation of this optimization problem is empirical risk minimization with ridge regularization.
Alternatively, this optimization problem can be interpreted as MAP estimation when assuming a Gaussian prior distribution $p_W(\mathbf{w}) \propto \exp(-\frac{\lambda}{2} \| \mathbf{w} \|_2^2)$ over the parameters $\mathbf{w}$ and that the observed data follows Gaussian noise. That is,
\begin{align}
\mathbf{w}_{\mathrm{MAP}} 
&= \arg\max_{\mathbf{w}} p_{W}(\mathbf{w} \mid \mathbf{y}, X) 
   = \arg\max_{\mathbf{w}} p_{Y}(\mathbf{y} \mid X, \mathbf{w}) p_W(\mathbf{w}) \label{eq:map1}\\
&= \arg\max_{\mathbf{w}} \ln p_{Y}(\mathbf{y} \mid X, \mathbf{w}) + \ln p_W(\mathbf{w}) \label{eq:map2}\\
&= \arg\min_{\mathbf{w}} \| X \mathbf{w} - \mathbf{y} \|_2^2 + \lambda \| \mathbf{w} \|_2^2 
   = \mathbf{w}^*. \label{eq:map3}
\end{align}
The key point here is that we have carefully selected the likelihood function and prior distribution such that the solutions to existing optimization problems correspond to MAP estimates.

Following the same strategy, we now show that iCNN can also be understood as MAP estimation. The optimization problem of iCNN is formulated as follows:
\begin{equation}
    \mathbf{s}^* = \arg\min_{\mathbf{s} \in \mathcal{M}} L(f(\mathbf{s}), \hat{\mathbf{\boldsymbol{\phi}}}), \label{eq:icnn-optimization}
\end{equation}
where $\mathcal{M} = \{ g(\mathbf{z}) \mid \mathbf{z} \in Z \}$ represents the set of images (generator manifold) generated by the pre-trained generative model $g: Z \to S$.
Here, $f: S \to \Phi$ is an encoder that extracts features from images, $\hat{\boldsymbol{\phi}} \in \Phi$ is a DNN feature predicted from brain activity, and $L: \Phi \times \Phi \to \mathbb{R}_{\geq 0}$ is a reconstruction loss function, typically squared error.
We instantiate the latent space and image space as $Z \simeq \mathbb{R}^{d_z}$ and $S \simeq \mathbb{R}^{d_s}$ (where $d_s \gg d_z$), and assume $g: \mathbb{R}^{d_z} \to \mathbb{R}^{d_s}$ to be a smooth injective function.
While neural generators with ReLU activations do not strictly satisfy these properties, this idealization serves our theoretical interpretation without affecting the actual implementation.
Let $J_g(\mathbf{z}) \in \mathbb{R}^{d_s \times d_z}$ denote the Jacobian of $g$, and define $G(\mathbf{z}) = J_g(\mathbf{z})^\top J_g(\mathbf{z}) \in \mathbb{R}^{d_z \times d_z}$.
We assume the following prior distribution and likelihood function for the latent variable $\mathbf{z}$ and DNN features, respectively:
\begin{align}
p_Z(\mathbf{z}) &\propto \sqrt{ | \det G(\mathbf{z}) | } \label{eq:prior-pullback} \\
    p_{\Phi}(\hat{\boldsymbol{\phi}} \mid \mathbf{z}) &\propto \exp(-L(f(g(\mathbf{z})), \hat{\boldsymbol{\phi}})) \label{eq:likelihood-reconstruction}.
\end{align}
The prior distribution in \autoref{eq:prior-pullback} has a specific geometric interpretation: it corresponds to pulling back a uniform distribution on the generator manifold $\mathcal{M}$ to the latent space.
To see why, suppose we assign a uniform (constant) prior $p_S(\mathbf{s}) = 1/|\mathcal{M}|$ only over images in $\mathcal{M}$ and zero elsewhere, where $|\mathcal{M}|$ denotes the volume of the manifold.
By the change of variables formula \autoref{eq:change-of-variables}, when we express this distribution in terms of the latent variable $\mathbf{z}$, we obtain $p_Z(\mathbf{z}) = p_S(g(\mathbf{z})) \cdot \sqrt{|\det G(\mathbf{z})|}$.
Since $p_S(g(\mathbf{z}))$ is constant for all $\mathbf{z}$, this yields $p_Z(\mathbf{z}) \propto \sqrt{|\det G(\mathbf{z})|}$ as stated.
Note that the change of variables applied here generalizes the standard formula to the case where $g$ is injective rather than bijective.
Thus, this prior effectively imposes no preference among images on the manifold while accounting for the geometric distortion introduced by the generator mapping.
Under the above probabilistic assumptions, MAP estimation on the manifold $\mathcal{M}$ spanned by the generator output becomes equivalent to the iCNN optimization problem.
We now demonstrate this equivalence step by step:
\begin{align}
\mathbf{s}_{\mathrm{MAP}}
&= \arg\max_{\mathbf{s} \in \mathcal{M}} p_S(\mathbf{s} \mid \hat{\boldsymbol{\phi}}) 
= \arg\max_{\mathbf{s} \in \mathcal{M}} p_{\Phi}(\hat{\boldsymbol{\phi}} \mid \mathbf{s}) \, p_S(\mathbf{s}) \label{MAP:eq1} \\
&= \arg\max_{\mathbf{s} \in \mathcal{M}} 
\ln p_{\Phi}(\hat{\boldsymbol{\phi}} \mid \mathbf{s}) 
+ \ln p_S(\mathbf{s}). \label{MAP:eq2}
\end{align}

To express the image prior $\ln p_S(\mathbf{s})$ in terms of the latent variable, we apply the change of variables formula \autoref{eq:change-of-variables}:
\begin{align}
    \mathbf{s}_{\mathrm{MAP}} &= \arg\max_{\mathbf{s} \in \mathcal{M}} \ln p_{\Phi}(\hat{\boldsymbol{\phi}} \mid \mathbf{s}) + \ln p_Z(g^{-1}(\mathbf{s})) - \ln \sqrt{\left| \det G(g^{-1}(\mathbf{s})) \right|} \label{eq:change-of-variables-with-G}.
\end{align}
Crucially, when we substitute our choice of prior from \autoref{eq:prior-pullback}, the Jacobian terms cancel exactly:
\begin{align}
    \mathbf{s}_{\mathrm{MAP}} &= \arg\max_{\mathbf{s} \in \mathcal{M}} \ln p_{\Phi}(\hat{\boldsymbol{\phi}} \mid \mathbf{s}) + \ln \sqrt{\left| \det G(g^{-1}(\mathbf{s})) \right|} - \ln \sqrt{\left| \det G(g^{-1}(\mathbf{s})) \right|} \\
    &= \arg\max_{\mathbf{s} \in \mathcal{M}} \ln p_{\Phi}(\hat{\boldsymbol{\phi}} \mid \mathbf{s}).
\end{align}
Finally, applying the likelihood definition from \autoref{eq:likelihood-reconstruction}, we recover the iCNN optimization:
\begin{align}
    \mathbf{s}_{\mathrm{MAP}} &= \arg\min_{\mathbf{s} \in \mathcal{M}} L(f(\mathbf{s}), \hat{\mathbf{\phi}}) = \mathbf{s}^*.
\end{align}
This transformation holds for all $\mathbf{s} \in \mathcal{M}$, but since $p_S(\mathbf{s})$ is not defined for $\mathbf{s} \notin \mathcal{M}$, the MAP estimation optimization is restricted to $\mathcal{M}$.
From the above discussion, we can see that the iCNN optimization problem can be interpreted as MAP estimation on the manifold spanned by the generative model.

This modeling reveals that deterministic methods like iCNN, which directly optimizes latent codes to minimize reconstruction loss $L(f(g(\mathbf{z})), \hat{\boldsymbol{\phi}})$, are mathematically equivalent to MAP estimation under the Jacobian-based prior described above.
While such probabilistic reinterpretations are theoretically valid, they offer no practical advantage unless the probabilistic framework is leveraged for enhanced inference capabilities such as uncertainty quantification, as discussed in the main text.

\end{document}